\documentstyle[12pt,epsfig]{article}

\begin{document}

\begin{center}

A critical examination of the spin dynamics in high-$T_C$ cuprates

\vspace{0.5 cm}
\noindent  
Ph. Bourges$^{1}$, B. Keimer$^{2}$,   
L.P. Regnault$^{3}$ and Y. Sidis$^{1}$

\end{center}

\vspace{0.5 cm}

\begin{tabular}{lp{5in}}
1& Laboratoire L\'eon Brillouin, CEA-CNRS, CE Saclay, 91191 Gif sur 
Yvette, France\\
2& Max-Planck-Institut f\"ur Festk\"orperforschung,
70569 Stuttgart, Germany\\
3& CEA Grenoble, D\'epartement de Recherche Fondamentale sur 
la mati\`ere Condens\'ee, 38054 Grenoble cedex 9, France\\
\end{tabular}

\vspace{0.5 cm}

\centerline{abstract}

A critical examination of the spin dynamics in high-$T_C$ cuprates is made 
on the light of recent inelastic neutron scattering results obtained 
by different groups. The neutron data show that incommensurate magnetic 
peaks in YBCO belong to the same excitation as the resonance peak observed 
at $(\pi/a,\pi/a)$. Being only observed in the superconducting state,
the incommensurability is then rather difficult 
to reconcile with a stripe picture. We also discuss the link 
between the resonance peak spectral weight and the superconducting 
condensation energy.

\vspace{1 cm}

After more than ten years of intense investigations, the precise
role of antiferromagnetic (AF) correlations for the mechanism of 
the high-temperature superconductivity remains a puzzling and open 
question. Since the early days, it has been obvious that both 
phenomena are clearly connected just by looking at the generic 
phase diagram of high-$T_C$ cuprates. Of course, a competitive 
role rather than cooperative between long-range antiferromagnetism 
and superconductivity was generally inferred as both 
phenomena are thought to occur in exclusion of each other. 
The next key question was then: are the dynamical AF correlations 
observed in the superconducting (SC) range of the phase diagram 
prejudicial or responsible for superconductivity ?

A necessary step to put some insight into this still unsolved
question is the knowledge of the spectral weight of the 
spin susceptibility, $\chi (Q,\hbar\omega)$. $\chi (Q,\hbar\omega)$
would, for instance, enter  the SC pairing interactions 
in any mechanism based on antiferromagnetism. As a matter of fact,
Inelastic Neutron Scattering (INS) is the only technique which 
directly measures the full energy and momentum 
dependences of the imaginary part of the spin susceptibility.
Further, the amplitude of $Im \chi (Q,\hbar\omega)$ can be 
determined in absolute units by a calibration of the 
magnetic neutron intensity versus other scattering such 
as phonons. This has been done only recently for the high-$T_C$ 
cuprates and, brings essential insight for the relation 
between AF correlations and superconductivity as we 
shall see below.

This technique is limited by the need of large 
single crystals (of cm$^3$ size) usually difficult to grow 
in complex systems such as high-$T_C$ cuprates. This has
reduced the number of systems which could be studied to a 
very few: $\rm La_{2-x} Sr_x CuO_4$ (LSCO), 
${\rm YBa_2Cu_3O_{6+x}}$(YBCO) and only recently 
 $\rm Bi_2 Sr_2 Ca Cu_2 O_{8+\delta}$ (BSCO). 
Further, INS spectra can be sometimes ambiguous to analyze
as, for instance, neutron scattering also directly measures 
the phonon spectrum which is typically of the same 
order of magnitude. Of course, a lot of effort has 
been developed to overcome these difficulties. However, 
this situation has postponed the emergence of a full agreement 
between the different groups. 
However, clear unmistakable features have been established 
which  have considerable impact on the role of AF fluctuations.
Here, we shall emphasize some key aspects on the basis 
of published data by the different groups. 
It should be mentioned that as far as the raw data are 
concerned, a fairly good agreement can be noticed. Disagreements
are rather related to the data analysis which sometimes leads to 
clearly different conclusions.

\vskip 1 cm
\noindent
{\bf  The resonance peak: a collective spin excitation of $d$-wave
superconductors}
\vskip .5 cm

Among the observed magnetic features \cite{revue-cargese}, 
the "AF resonance peak" 
observed below $T_C$ is certainly one of most important 
results which has been widely studied since its discovery in 
1991 by Rossat-Mignod {\it et al} \cite{rossat91} in YBCO$_{6.92}$.
When entering  the SC state and {\bf only} below $T_C$, a 
sharp (almost energy resolution limited) spin excitation appears in the 
neutron scattering data at an energy, $E_r$ and at the AF wavevector 
$Q_{AF}=(\pi/a,\pi/a)$ ($Q_{AF}$ is the propagation wavevector of the 
AF state of the insulating undoped parent compound, a=3.85 \AA\ 
is the 2D square lattice parameter).

The striking characteristic of the resonance peak is actually its 
temperature dependence. Indeed, its energy, $E_r$, does not shift towards 
lower energy when approaching $T_C$ (a shift of at most $\sim$ 4 \%  can be 
inferred \cite{hoechst,tony2,dai96}) but its intensity is vanishing 
upon heating at the superconducting temperature $T_C$ for all doping levels, actually following an order parameter-like behavior. 
Recently, an attempt has been made\cite{dai99} to associate
the vanishing of the resonance peak intensity in underdoped sample 
with the temperature T$^*$ where the resistivity displays the so-called 
"pseudo-gap" anomaly. This statement is not correct being based on 
an arbitrary analysis. Indeed, neither the data published by 
Dai {\it et al} \cite{dai96,dai99} nor our own data 
\cite{epl,tony3,tony2000} provide any justification for a separation in 
the normal state (NS) of the spectrum into resonant and nonresonant parts. 
No published temperature dependence of the neutron intensity at 
the resonance energy  (or more correctly, at  the 
energy transfer where the resonance peak appears in the SC state) 
suggests an anomaly at a temperature T$^*$ larger than $T_C$.
A clear upturn is systematically observed only at the SC transition 
temperature. In our opinion, this incorrect attribution of the 
''onset of the resonance peak at T$^*$'' has been made from the 
fact that the broad maximum of the spin 
susceptibility in the normal state occurs in some underdoped sample roughly 
at the same energy as the resonance peak \cite{revue-cargese,tony2000}. 
But, as a matter of fact, the apparent equivalence of the normal state 
energy and the resonance peak energy breaks down in underdoped samples closer 
to optimal doping \cite{revue-cargese,miami}. 

Interestingly, the resonance energy 
scales with the SC temperature as: $E_r \sim 5.2 k_B T_C$.
This relation holds in the two systems where the resonance peak 
has been observed so far, YBCO\cite{revue-cargese,tony2000} 
and BSCO\cite{bi2212}. This actually is not only valid at optimal doping 
but also remains correct on both sides of the high $T_C$ phase diagram: 
on the underdoped side, as experimentally realized for different oxygen 
contents in YBCO\cite{revue-cargese,tony2000}, as well as on 
the overdoped side as observed in a BSCO sample\cite{he}. This generic 
relationship of $T_C$ with the temperature-independent resonance energy 
calls for an explanation which is not obvious when one considers the 
different models usually invoked to interpret the resonance peak
(See Refs \cite{revue-cargese,tony2000,he} for a discussion of these
approaches).

Further, the resonance feature appears to be strongly sensitive to 
parameters which affect the superconducting properties. For instance, 
the substitution of Zn impurities within the CuO$_2$ plane in YBCO, known 
to strongly reduce the SC temperature ($dT_C/dy \simeq -12$ K/\%) \cite{zntc}
without changing the doping level\cite{alloul}, 
induces a rapid vanishing of the resonance intensity: small amounts 
of zinc impurities ($y$ ranging from 0.5\% to 2\% in 
YBa$_2$(Cu$_{1-y}$Zn$_y$)$_3$O$_{6+x}$)\cite{sidis96,tonyzn,yvani} are 
sufficient to remove its spectral weight without strong 
renormalization of the resonance energy itself. In contrast, 
magnetic Ni impurities which are three times less efficient to 
remove superconductivity ($dT_C/dy \simeq -4$ K/\%) \cite{zntc}, 
have also less effect on the resonance peak intensity and keep
 the ratio $E_r/k_B T_C$ almost unchanged \cite{yvani}. This extreme
sensitivity of the resonance feature to defects affecting the SC 
transition
temperature then might explain why no resonance peak has been reported 
so far in the LSCO system whose maximum $T_C$ ($\sim$ 40 K) is anomalously 
low as compared to other single CuO$_2$ layer systems where 
$T_C$ can reach 90 K (Tl- or Hg- based system). The disorder which might 
be responsible for the reduction of $T_C$ in LSCO can also remove 
the resonance peak feature. Further, Zn and Ni impurities in 
YBCO\cite{tonyzn,yvani} also produce a systematic 
broadening in energy of the resonance peak, by $\sim$ 10 meV. Similar 
broadening found in BSCO \cite{bi2212,he} can then be naturally 
accounted for by the presence of intrinsic defects in that system.

\begin{figure}[tbp]
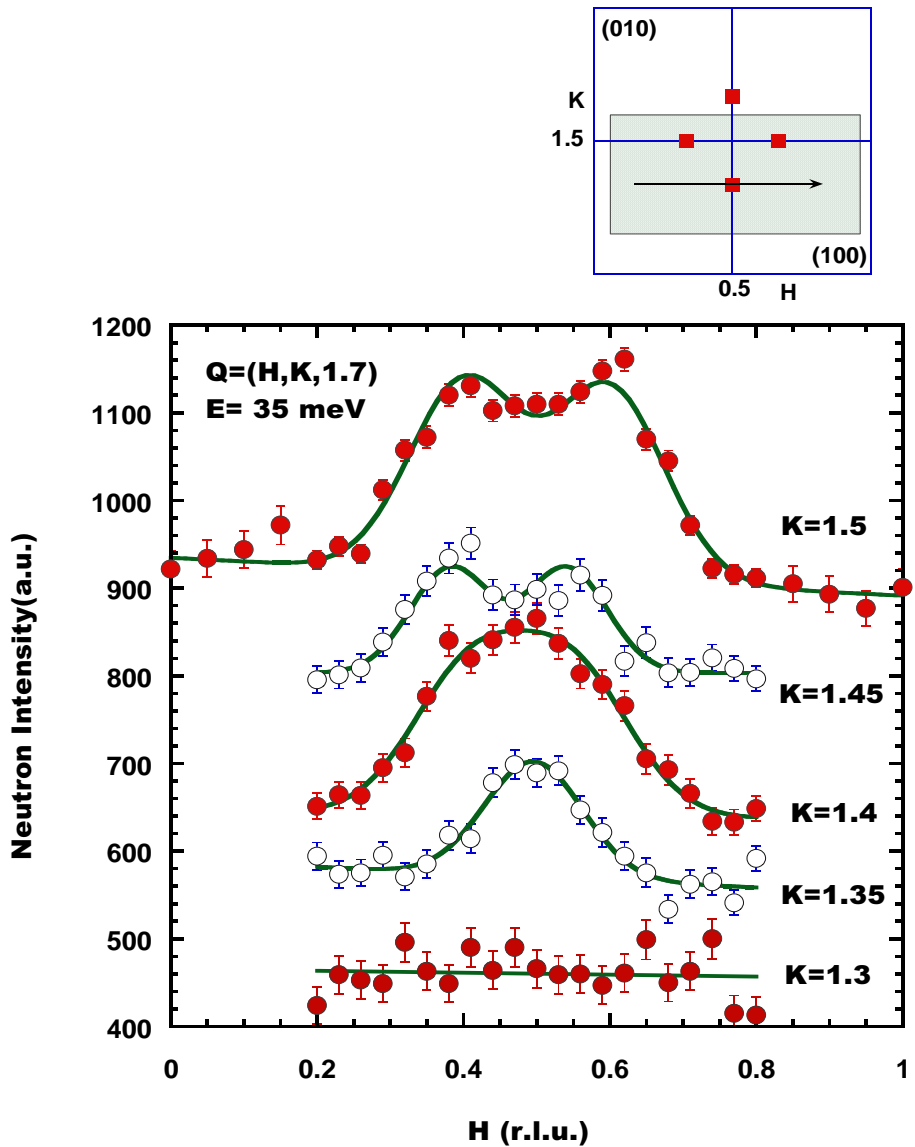

\caption{ Lower panel) Neutron intensity scans in the $(H,K)$ momentum plane 
for a fixed energy transfer at E= 35 meV (each scan has been 
shifted by 120 cnts from the lower one for presentation). The momentum 
transfer along $c^*$ was fixed to the maximum of the magnetic 
structure factor $L=1.7$\cite{tony2000}. 
The phonon background measured 
at room temperature has been subtracted from the data after proper 
correction of the temperature factor following a procedure detailed 
in \cite{epl}. Measurements have been performed on the 2T triple-axis 
spectrometer (Laboratoire L\'eon Brillouin, Saclay) with 
$k_f$=2.662 \AA$^{-1}$, the momentum resolution (FWHM) was 0.14 r.l.u. 
along $H$ direction and 0.1 r.l.u. along  $K$ and the energy 
resolution was 4 meV. All scans have been fitted by either two Gaussians 
peaks displaced by $\Delta H$ from  $H=0.5$ or a single Gaussian 
peak centered at $H=0.5$.
Upper panel) Sketch of the reciprocal space around the AF wavevector.
The squares represent the locus of maximum magnetic intensity in the 
superconducting state. The shaded area indicates the momentum 
$(H,K)$ space covered by the q-scans of the lower panel.
}
\label{pattern}
\end{figure}

Until recently, the resonance peak has been widely described as a 
single commensurate excitation. Although this statement remains 
certainly correct in the slightly overdoped YBCO$_7$ system, we have recently 
demonstrated\cite{science} in YBCO$_{6.85}$ that the 
resonance peak actually exhibits a full dispersion curve away from 
$(\pi/a,\pi/a)$ momentum. This illustrates, on experimental grounds, that 
the resonance peak can be considered as a collective mode of the 
superconducting state of high-$T_C$ cuprates as 
theoretically proposed (see e.g. \cite{vdM,OP,BL}). 
The observed downward dispersion actually relates the 
commensurate resonance peak with the incommensurate peaks observed 
at lower energy and recently reported in underdoped YBCO\cite{mookinc,arai}.
By detailed temperature dependences of the neutron intensity at 
different wavevectors and energies, we have established\cite{science} 
a dispersion compatible with the following relationship,

\begin{equation}
E_r(q)=\sqrt{E_r^2(Q_{AF})-(\alpha q)^2}
\label{dispersion}
\end{equation}
where $q$ is the wavevector measured from $Q_{AF}=(\pi/a,\pi/a)$.
$E_r(Q_{AF})=41$ meV is the previous commensurate resonance energy, and
$\alpha \simeq$ 125 meV.\AA\ represents an isotropic dispersion relation. 
Certainly, the relation Eq. \ref{dispersion} is only a first 
approximation which needs to be refined. Indeed,
the measured wavevector pattern at a fixed energy E= 35 meV 
located below $E_r(Q_{AF})$ exhibits an intensity modulation in 
the 2D $(H,K)$ momentum space shown in Fig. \ref{pattern} with larger
intensity in the directions (100) or (010) and lower 
intensity in the directions (110) or (1$\overline{1}$0). 
Such detailed momentum dependence (which reproduces the shape 
reported in YBCO$_{6.6}$\cite{mookinc} at 24.5 meV (below 
$E_r(Q_{AF})=34 $ meV) as well as that discussed in \cite{arai}), 
implies a modification in the dispersion relation  of Eq. \ref{dispersion}.
For instance, an anisotropy of $\alpha$ between the (100) and (110) 
directions should be added and would certainly account for 
the momentum pattern of the neutron intensity shown in Fig. \ref{pattern}.
Although the resonance peak dispersion is, so far, only evidenced in one 
sample, YBCO$_{6.85}$, we think it is a generic feature of the spin 
dynamics in the superconducting state over a wide part of the high-$T_C$ 
cuprate phase diagram. Data reported in Refs. \cite{mookinc,arai} are 
fully consistent with such an interpretation although this has not 
been discussed this way. 
For sure, more work is needed to generalize this conclusion,
for instance, to give the actual doping dependence of the $\alpha$
parameter. 

The observation of incommensurate peaks \cite{mookinc,arai}, 
in addition to the commensurate resonance peak, has stimulated several 
theoretical models in Fermi liquid-like theories. It has been discussed 
as a combined effect of both i) topology of the band structure 
and ii) anisotropic  superconducting  order parameter either at 
the level of the bare susceptibility \cite{VooandWu,abrikosov} or 
after taking into account of the interactions by a
random phase approximation \cite{BL,levin,norman}. 
Furthermore, a dispersive collective mode has been predicted 
to arise below the particle-hole spin-flip continuum in the $d$-wave
superconducting state as a result of a momentum-dependent pole in the 
spin susceptibility pulled by antiferromagnetic interactions \cite{OP}.
Our recent observation of a downward dispersion \cite{science}
supports the latter proposal. However, to fully establish the collective 
nature of the resonance peak, a necessary step will be to 
observe the particle-hole spin-flip continuum. 
In any case, our recent data demonstrate that superconductivity affects 
not only the energy lineshape of the spin susceptibility by 
inducing a resonance peak at $(\pi/a,\pi/a)$ but also that
it drastically changes its momentum dependences.

\vskip 1 cm
\noindent
{\bf  ''Incommensurate peaks'' in YBCO: not an evidence for dynamical stripes}
\vskip .5 cm

The observation of ''incommensurate peaks'' at some energy transfers 
\cite{mookinc,arai} has often been interpreted as  clearcut 
evidence of dynamical stripes in YBCO. Our recent detailed 
study\cite{science} basically rules out this conclusion
(at least for near-optimally doped YBCO). Indeed, 
we established that the ''incommensurate peaks'' are only observed in 
the superconducting state and are additionally closely related to the 
commensurate resonance peak by a continuous  dispersion relation 
(Eq. \ref{dispersion}) as discussed above. This puts the observation 
of the magnetic incommensurability in  YBCO  in a totally new perspective.  
Being energy-dependent, temperature-dependent and doping-independent, 
the ''discommensuration'' is rather difficult to understand within 
a stripe picture where typically a characteristic distance (between 
charge stripes) needs to be observed. Without invoking any specific model, 
it becomes clear that their interpretation has to be necessarily related to 
the one made for the ''commensurate'' resonance peak. 

The vanishing at $T_C$ of the ``incommensurate'' excitations, 
we reported in YBCO$_{6.85}$ \cite{science}, can be actually anticipated 
over a wide part of the phase diagram\cite{miami}. 
[Notice that, even below $T_C$, it 
is still not established under which conditions and exactly in which 
doping range the ``incommensurate'' excitations are present in YBCO.] 
Nevertheless, their disappearance in the normal state is actually consistent 
with the different data published so far\cite{mookinc,arai,incdai}. 
Indeed, the reports of normal state incommensurability in YBCO are rather 
scarce. At best, it is said that these incommensurate excitations remain
in a small temperature window above $T_C$ (up to 70-75 K for 
$T_C$=63 K) \cite{incdai} and finally disappear 
upon heating. But, as this intensity is weak on top of a phononic 
background (always present in such unpolarized neutron scattering 
experiments) whose structure factor mimics an incommensurate-like 
intensity modulation, no clear conclusion can be made and, at least, 
requires further work. In any case, fluctuations of the 
SC state (in the conventional meaning) could also explain the persistence
of ``incommensurate'' excitations in a small temperature range above $T_C$.  
Recently, it has been argued that these  incommensurate  magnetic 
fluctuations have a one-dimensional nature \cite{detwinned}.
This is based on measurements using a partially (half) detwinned 
YBCO$_{6.6}$ sample. Due to the above-mentioned phononic background
and the scattering geometry used, this report is rather inconclusive: 
it is not proved that the observed effect is related to the magnetic 
scattering. Indeed, the detwinning of the sample can actually affect 
the background itself (for instance, if it is related to an 
$a^*$-polarized phonon). To make their point clear, these authors have 
to demonstrate that the balance of intensity between $a^*$ and $b^*$ is not 
present at high temperature (where the magnetic intensity is weaker 
and commensurate) or present polarized neutron beam data.

Our results in YBCO$_{6.85}$ \cite{science} also contrast with those
reported in the LSCO system \cite{aeppli,lake} where no change of
the incommensurate peak position occurs across the superconducting 
temperature. In LSCO, the incommensurate peak structure begins to 
disappear only around room temperature \cite{aeppli}. Further, the observed 
energy range where incommensurate peaks are observed is very different 
in the two systems (down to the lowest energies in LSCO but limited 
in a small energy range below $E_r$ in YBCO). Their similarity is 
then reduced to only the symmetry 
of the incommensurate pattern along the (100) or (010) directions seen 
in both systems. However, the actual ``fortified castle''-like shape 
observed in YBCO\cite{mookinc} looks rather different from the four 
well defined peaks observed in LSCO\cite{lake}. This makes dubious 
the universality of the spin fluctuations claimed to occur in the 
two systems\cite{mookinc} only based on the occurrence of ``incommensurate'' 
magnetic peaks. The origin of incommensurability in both systems 
likely requires a different scenario although common ingredients (such as 
Fermi surface topology) might be invoked.

As discussed above, our detailed study of the incommensurate magnetic 
peaks in YBCO shows that a standard interpretation within a 'stripe phase' 
picture is inconsistent. However, it should be noticed that a situation 
of strongly disordered stripes, as recently theoretically discussed in 
\cite{cneto}, is still possible. This would correspond to the case where 
the AF correlation length is lower than the mean distance between stripes 
\cite{tranquada}. And so, there is no $\pi$-phase shift from one AF 
cluster to the next one. These decorrelated AF clusters would give rise to 
the broad commensurate peaks observed in the normal state. However, 
the behavior of such objects in the superconducting state has not been 
addressed so far. This would be of great interest.

\vskip 1 cm
\noindent
{\bf  Resonance peak and Superconducting condensation energy}
\vskip .5 cm

The knowledge of the spin-spin correlation function in absolute units 
is becoming a crucial topic for the description of the physical 
properties of high-$T_C$ cuprates. For instance, magnetic neutron scattering
has been recently proposed to provide a direct measurement
of the condensate fraction of a superconductor\cite{sudip}.
A direct link with the high-$T_C$ mechanism has also 
addressed in the framework of the t-J model\cite{scalapino,demler}.
The proposal is the following: if the SC pairing mechanism is due to 
AF exchange then the SC condensation energy, $E_C$, would be the energy
gain between the normal state and the superconducting state of an exchange 
energy $E_J$ of the form\cite{scalapino,demler}:

\begin{equation}
 E_J =  \frac{3 J}{2 \pi (g \mu_B)^2} 
\int_{BZ} d^2q [\cos (q_x a) + \cos (q_y a)]
\int d\omega \frac{ Im \chi (q,\omega) }
{1 - \exp(-\hbar\omega/k_B T)} 
\label{EJ}
\end{equation}
where the 
sum over the wavevector is performed over the 2D Brillouin zone (BZ)
and normalized by the BZ volume, $(2 \pi/a)^2$.
The condensation energy then reads, 
\begin{equation}
 E_C =  E_J^{NS}- E_J^{SC} 
\label{dEJ}
\end{equation} 

It is essential to realize that Eq. \ref{dEJ} is a subtle net difference 
of the magnetic fluctuations spectral weight between the normal state
and the superconducting state additionally {\bf weighted}  by a momentum 
form factor $[\cos (q_x a) + \cos (q_y a)]$ corresponding to the 
Fourier transform of the AF exchange. It follows that the 
temperature dependent change in exchange energy 
crucially depends on a redistribution of the magnetic spectral weight in 
{\bf momentum}. Indeed, according to Ref. \cite{scalapino}
the exchange energy differs from the total moment sum rule,
$W=\int_{BZ} d^2q d\omega Im \chi(q,\omega)/(1 - \exp(-\hbar\omega/k_B T))$,
only by this momentum-dependent form factor. If one  neglects this 
wavevector dependence in Eq. \ref{EJ}, Eq. \ref{dEJ} becomes 
meaningless as $E_C$ will necessarily be zero to satisfy the sum-rule.
 The wavevector form factor in Eq. \ref{EJ} is then
essential and cannot be neglected. 

In a recent Report, Dai et al. \cite{dai99} have followed this idea 
and claim to have found a quantitative correspondence between the 
temperature derivative of the spectral weight of spin excitations in 
YBCO and the electronic specific heat $\rm C_{el} \simeq d E_J/d T$. 
We wish to point out that the analysis provided by Dai et al. fails at 
an elementary level as they fully neglected the wavevector form factor 
in Eq. \ref{EJ} by rewriting $E_J$ as,

\begin{equation}
 E_J \simeq  \frac{3 J}{ \pi (g \mu_B)^2} 
\int_{BZ} d^2q 
 d\omega \frac{ Im \chi^{res} (q,\omega) }
{1 - \exp(-\hbar\omega/k_B T)}
\label{EJdai}
\end{equation} 
where  $Im \chi^{res} (q,\omega)$ is only the resonant
part of the spin excitations. 
Eq. \ref{EJdai} is derived by assuming that the spins accounting 
for the resonance part are fully decorrelated in the normal state in  
contrast with the observation of AF dynamical correlations above 
$T_C$. They then conclude that a large part of the electronic specific heat is 
due to spin fluctuations. There is no doubt that the electronic specific 
heat and the spin fluctuations are related in some way: after all, they 
are ultimately attributable to the same strongly interacting electron 
system. However, the analysis of Dai et al. \cite{dai99} is much 
too crude to uncover this underlying relation. In an optimally doped
sample, they finally obtain a contribution to the specific heat 
$\sim$  {\bf three} times larger than the measured one\cite{loram}
(as found in \cite{demler}). In underdoped samples, 
the discrepancy is even bigger as the measured specific heat jump drastically 
falls down  whereas the resonance peak spectral weight remains 
approximately constant for all doping\cite{dai99,tony2000} as
$\int d^2q d\omega Im \chi^{res} (q,\omega) \sim 0.05 \pm .02\ \mu_B^2$, 
and so would be the calculated specific heat jump at $T_C$. 
 [It should be noticed that this absolute unit value 
has been independently obtained by the two different groups].
Further, the attempt to relate the specific heat anomaly in the normal 
state with a speculated onset of the resonance peak at $T^*$ (see above) 
is meaningless. Indeed, the most salient feature of the electronic specific 
heat \cite{loram} is its pronounced {\bf increase} with increasing 
doping in the normal state. By contrast, the magnetic spectral weight 
strongly {\bf decreases} with increasing 
doping in the same temperature range. 
These discrepancies do not necessarily suggest that the proposal
of Eq. \ref{EJ} is not correct. It just means that the analysis performed 
in Ref.\cite{dai99}, Eq. \ref{EJdai}, relating the magnetic 
fluctuation spectrum and the  electronic specific heat is invalid 
and inconclusive as it oversimplifies the physical content of Eqs. \ref{EJ}
and \ref{dEJ}. 
As emphasized by Scalapino and White \cite{scalapino}, the net difference 
in Eq. \ref{dEJ} will be very small and then difficult to estimate. 
To overcome this problem, Dai et al. \cite{dai99} have arbitrarily 
considered {\bf only } the contribution of the resonance peak spectral weight 
around $(\pi/a,\pi/a)$ and at the energy $E_r(Q_{AF})$ 
(that they attempt to relate to the electronic specific heat).
The actual change of the spin susceptibility across the SC temperature 
as discussed above (dispersion behavior such as Eq. \ref{dispersion}) 
reveals that the estimate of Eq. \ref{dEJ} would be very subtle 
(especially in underdoped samples).
 
In conclusion, the resonance peak is certainly a key feature for the 
description of the physical properties of high-$T_C$ superconductors
which has been widely reported at the commensurate AF wave vector.
Further, the observation of its dispersion\cite{science} experimentally 
suggests its collective nature. It now emerges that the role of such 
a magnetic collective mode would be essential for the interpretation
of physical properties of high-$T_C$ superconductors, 
for instance, to describe
the complex spectral structure of the one-particle spectrum\cite{campuzano} 
as reported by photoemission spectroscopy. 

\vskip 0.4 cm

Acknowledgments:

We wish to thank A.H. Castro Neto, G. Deutscher, D. Pavuna 
for stimulating discussions at the Klosters conference.


 \clearpage

\vskip 1 cm \ \ \
\epsfig{file=klo_fig.epsi,height=15 cm,width=12 cm}

\end{document}